\documentstyle[11pt, aaspp4]{article}

\begin{document}

\title{The Energy Dependence of the Aperiodic Variability \\
for Cygnus X-1, GX 339-4,
 GRS 1758-258 \& 1E 1740.7-2942}
\author{ D. Lin\altaffilmark{1}, I. A. Smith\altaffilmark{1}, 
M. B\"ottcher\altaffilmark{1, 2}, 
\& E. P. Liang\altaffilmark{1}}

\altaffiltext{1}{Department of Space Physics and Astronomy, 
Rice University, 6100 S. Main, Houston, TX 77005, USA. 
Email: lin@spacsun.rice.edu.}

\altaffiltext{2}{Chandra Fellow.}

\begin{abstract}
Using the data from the {\it Rossi X-ray Timing Explorer (RXTE)}, 
we report the different energy dependence of the variability of 
the four persistent hard X-ray sources in the low-hard state: 
Cygnus X--1, GX 339--4, GRS 1758-258 and 1E 1740.7--2942. 
Cygnus X--1 is found to have a flatter power density spectrum (PDS)
shape at higher energies. The other three sources have energy 
independent PDS shapes. The energy dependence of the overall 
variability (the integrated rms amplitude) varies from source
to source and from observation to observation. 1E~1740.7-2942, 
for example,  has a variability generally increasing with energy
while GX~339-4 has a decreasing variability. A general trend
is found in the four sources that the integrated rms amplitude 
anti-correlates with the X-ray flux. We compare these distinct 
energy dependent behaviors with several emission models. 
None of the models can fully explain all the features that we 
have found.
\end{abstract}

\keywords{accretion, accretion disks --- black hole physics 
--- stars: individual (Cygnus X--1, GX 339--4, GRS 1758--258, 
1E 1740.7-2942) --- X-rays: stars}

\section{Introduction}

The galactic black-hole candidates (GBHCs) are known to exhibit
rapid variability on time scales of hours down to milliseconds.
This rapid aperiodic and quasiperiodic variability provides a 
diagnostic tool for understanding the nature of the high-energy 
emission in these objects (for reviews see \cite{van95};
\cite{cui98}; \cite{liang98}). The power density spectrum (PDS) 
of the variation is one of the widely studied properties. Generally 
the PDS of GBHCs consists of a flat plateau at low frequencies 
(red noise) up to a break frequency $f_1$ and a power-law, 
typically $\propto f^{-1}$ for $f > f_1$. In several sources, a 
second break is observed at frequency $f_2$, beyond which 
the PDS turns into a steeper power-law, typically $\propto 
f^{-\beta}$ with $1.5 \lesssim \beta \lesssim 2$. Hereafter, 
we will refer to this structure using a double broken power-law 
model (although it can generally be equally well described 
using two zero centered lorentzians). Superposed on the above 
shape  some GBHCs show ``peaked noise'' or QPOs in the 
power-law section of the PDS. The QPOs, which may be related 
to the dynamic processes in the accretion disks, have received 
widespread attention. In this paper, however, we focus on the 
energy dependence of the overall PDS. 

Several X-ray emission models of the GBHCs have suggested
that the PDS should be energy dependent.  \cite{hua97} have 
proposed that Comptonization in an extended hot, thermal 
corona may be responsible for the PDS shape and the hard 
time lags of GBHCs. B\"ottcher \& Liang (1998) studied the 
pure Comptonization model in great detail for different source 
geometries. They assumed that the X-ray emission is generated 
by the upscattering of soft photons, which are injected from 
either a central or an external source. They showed that the 
Comptonization model with a central soft photon source 
generally has a steeper PDS at higher energies, but this steepening
trend is less significant when the density gradient of the
Comptonizing cloud increases. However,  if the soft photon 
source is external, the steepening trend is less significant 
with lower density gradient. To avoid the unrealistic
requirement of the pure Comptonization model for the huge 
size (e.g. $10^3$ -- $10^4$ $R_g$) of the Comptonizing corona,  
B\"ottcher \& Liang (1999) proposed that the soft photons may 
be injected from a cool blob spiraling into the central object. 
In this scenario, the simulation shows that the PDS becomes flatter
with increasing energies. In the magnetic flare  model of \cite{pf99},
a pulse-avalanche model is invoked in order to reproduce the overall
shape of the observed PDS of Cygnus X-1. In this model,
the general shape does not change with energy except that 
the second break frequency is shifted to lower frequencies for 
higher energy bands (Poutanen \& Fabian 1999; J. Poutanen 1999,
personal communication).

Given these different model predictions, it is important to find 
out in the available data how the PDS shape and integrated 
rms amplitude evolve with energy. The flattening PDS, as 
suggested by the spiraling blob model, was observed in 
Cygnus X--1 (\cite{cui97}; \cite{nowak98}). \cite{sl99} reported 
that there were no significant PDS shape changes among the 
energy bands for an {\it RXTE} observation of GX 339--4, and 
that the rms amplitude decreased with increasing photon energy. 
GRS 1758--258, on the other hand, had the highest rms amplitude 
in the medium energy band (\cite{lin99}). 
These different behaviors may be intrinsic to the sources, and 
thus different emission mechanisms or viewing geometries 
would be necessary for different GBHCs in spite of their spectral 
similarities. To address this issue, we have systematically 
analyzed the {\it RXTE} archival data for the four hard X-ray 
sources in their hard/low state: Cygnus X--1, GX 339--4, 
GRS 1758--258 and 1E 1740.7--2942. In section 2, we discuss 
how we select  and reduce the data. In section 3, we show the 
analysis results. In section 4, we compare the results to various 
model predictions. In section 5, we discuss the results.

\section{Data Selection and Reduction}

The data have been properly selected from the $RXTE$ archive 
over the period of 1996 -- 1997. We chose only those data that 
came from long enough observations and had proper data formats 
to allow us to do an accurate energy dependent analysis. 
In all cases, the sources were in the hard/low state,
which was determined by careful spectral analyses.
The dates of the observations are listed in Table 1.
In order to study the energy dependence of the variability, we divided
the PCA effective energy range into three energy bands, the ranges of
which are approximately: 2 -- 5 keV (L), 5 -- 10 keV (M), and 
10 -- 40 keV (H). Some of the archival data formats make the energy 
ranges a little different from these values. The counts have been 
summed into time bins of 46.875 ms, and the power density spectra 
generated on intervals of 384 seconds. We subtracted from the data 
the background count rate estimated from the  model L7/240 or 
(Q6+activation+X-ray). The contributions from the galactic diffuse 
emission, whose contamination to the X-ray flux of GRS 1758-258 
and 1E 1740.7--2942 is significant, have also been subtracted from 
the light curves of the two sources, as described in Lin at al. (1999); 
see also \cite{main99}. 

The power density spectrum for each band was generated using the FTOOLS
analysis package with the white noise subtracted. For comparative purposes,
we calculated the PDS over the frequency range of 0.002 -- 10 Hz for
all the sources even though some of them could be analyzed up to higher 
frequencies. We checked that we did not introduce any artificial effects 
by limiting the frequency range. For each observation, the final power 
density spectra have been averaged over all the observation segments. 
We calculated the H/L and M/L band ratios to investigate how the PDS
shapes change with energy (the error of the PDS ratio is the propagating
error). For each band we also integrated the PDS over the frequency 
range of 0.002 -- 10 Hz, and took the square root to calculate the overall 
integrated rms amplitudes.

\section{Analysis Results}

\subsection{Power Spectrum Shape}

In Figure 1, we show an example of the energy dependent PDS and 
their ratios for one observation of Cygnus X-1. Similar results are seen in all
seven observations of Cygnus X-1.  The PDS ratios between the
high- and low- energy bands significantly increase above 3 Hz.
Below 3 Hz, the PDS ratios are consistent with  constant values
though there appears to be a slowly increasing trend. Moreover, 
we found that the two PDS ratios ${\rm PDS_{H}/PDS_{L}}$ and 
${\rm PDS_{M}/PDS_{L}}$ start to increase roughly at the same 
frequency (3 Hz). Therefore, in terms of the double broken power 
law model, $f_1$ and $f_2$ are energy independent and $\beta$ 
decreases with energy. The decrease in $\beta$ means that the 
PDS shape becomes flatter at higher energies above $f_2$. 
This flattening trend in Cygnus X-1 has been previously reported 
(\cite{van95}; \cite{cui97}; \cite{nowak98}). The flat distribution of 
the PDS ratio over the frequency range between $f_{1}$ and $f_{2}$ 
indicates that the first power-law index, which is usually close to 
1, indeed has little energy dependence.

No significant trends are seen for the other three sources.
In all six observations of GX 339--4, we found the PDS ratios
to be consistent with constants over the frequency range of 
0.002 -- 10 Hz (e.g. Figure 2), and we did not see any significant 
increases beyond 10 Hz either. Constant PDS ratios mean that the 
PDS shapes are energy independent, and thus $f_1$, $f_2$ and $\beta$ of
the double broken power law model have little energy dependence.
Constant ratios were also found in all six observations of GRS 1758--258 
(e.g. Figure 3) although detailed model fittings indicate that 
QPOs in the PDS are energy dependent (\cite{lin99}). In three 
out of the five observations of 1E 1740.7--2942 the PDS ratios
are also consistent with constants
(e.g. Figure 4). In the other two observations of 1E 1740.7--2942, 
we found there were several big flares in the light curves. These
flares, which have significant energy dependence, may be due to 
the nearby burst source previously detected by $RXTE$ (\cite{strohmayer97}). 
The burst source is located $0.8^{\circ}$ away from the PCA pointing position.
By excluding these flares from the timing analysis, we found
the PDS ratios in these two observations are also consistent with a 
constant over 0.002 -- 10 Hz. A complete analysis of the flares found 
in our 1997 Nov 11--12 observation of 1E 1740.7--2942 will be 
described in a separate paper (\cite{lin00}) along with a full spectral 
analysis of 1E 1740.7--2942.

\subsection{Overall Variability}

The behaviors of the overall variability, represented by the integrated 
rms amplitude over the frequency range of 0.002 -- 10 Hz,  are significantly 
different among the four sources (Figure 5). 1E~1740.7--2942 has the 
lowest variability in the low energy band and has a general trend of increasing 
variability with energy except for the observation made in 1996 (MJD 50158), 
which has the highest variability in the medium band. 
GRS~1758--258 has  the highest variability in 
the medium energy band and the lowest variability in the high band.  
GX~339--4 has a generally decreasing variability
with increasing energy, except for two observations in which the medium energy
band has slightly higher variability than the low band.
Among the seven observations of Cygnus X--1, we found four cases with
decreasing variability, one with constant variability and two with increasing
variability, and the pattern appears to depend on the overall variability
of the source.

Previously Cygnus X--1 and GX 339--4 have been reported to have little
energy dependence (\cite{van95}). This does not contradict our results.
We have one observation of Cygnus X--1 (on MJD 50725) in which the 
overall variability is energy independent, and two other observations in 
which the integrated rms amplitude changes by less than 0.012 from the low
to high energy bands. Similarly, in one observation of GX 339--4 (on MJD 50636),
we see a change  of less than 0.028 in the integrated rms amplitude
over the three energy bands. Such small effects would have been  indistinguishable in 
the early measurements (e.g. \cite{nolan81}; \cite{maejima84}). 

When one considers that the spectra of the four sources are similar in
the hard/low state, it seems surprising that the overall variability behaviors
are so different. It is interesting to note that only one spectral property,
the hydrogen column density due to the ISM and any materials local to the source
is significantly different among the sources. The typical hydrogen 
column density for each source is shown in Figure 5. We see a tendency 
that the hydrogen column density anti-correlates with the variability 
of the source, especially in the low energy band.
One possible effect that the absorption has on the X-ray variation is that it 
might diminish the variability of the X-ray flux by absorbing and 
scattering the X-ray photons. 1E 1740.7--2942, for example,  has the 
highest value for $N_{H}$, thus the diminishing effects on its X-ray flux can be 
significant up to 10 keV. This could make the lower two bands less variable
if, for example, bright flares emitted from one region in the system are more 
absorbed than the general persistent emission.

By comparing the rms amplitudes with the related $RXTE$ all sky monitor (ASM) 
(2 -- 12 keV) count rates, we found a tendency towards an anti-correlation between
the overall variability and the X-ray flux (Table 1 and Figure 6). Due to closeness
of their values and the big statistical errors, the rms amplitudes 
and ASM count rates of the observations on MJD 50470 and 50517 for
GRS 1758-258 have been combined, as have the observations on MJD 50158--61
and 50764 for 1E 1740.7-2942. This anti-correlation is consistent with previous 
timing results of Cygnus X--1, GX 339--4 and GS 2033+38 (\cite{van95}; 
\cite{nolan81}; \cite{miyamoto92}), and can also be identified in the long-term 
monitoring of 1E 1740.7--2942 and GRS 1758--258 (see Figure 2 in Main et al. 1999).
Again, this anti-correlation may imply that flaring and non-flaring
emission regions are physically separated. With the increase of
the non-flaring emission, the relative variation in the flux would
decrease.  We also compared the rms amplitude with the related 
BATSE Earth Occultation flux (20 -- 120 keV),  but no significant 
correlations were found.

\section{Comparison with Aperiodic Variation Models}

Since the discovery of rapid aperiodic X-ray intensity variations in 
Cygnus X--1 by \cite{oda71}, various
models of the aperiodic variations have been proposed. The early
models were mostly phenomenological, such as shot models,
but several more physical models have recently been proposed.

\subsection{Shot Models}

The shot models, originally proposed by \cite{terrell72}, are good at mimicking
the observed light curves by superposing individual shots
with  properly chosen shot profiles. However, it is not clear 
how the shots are generated in the accretion disk (\cite{van95}; \cite{cui98}). 
\cite{takeuchi95} proposed a shot generation mechanism based on 
self-organized criticality (SOC). In the SOC model, the mass 
accretion takes place in the form of avalanches
only when the local density exceeds some critical value. The model 
can produce the 1/f-like power density spectrum and predicts
a positive correlation between the dissipated energy and the duration
of the shot. The photon energy in each shot is determined by the local 
temperature of the disk where the shot is generated. Therefore,
the SOC model alone does not make any predictions about the 
energy dependence of the PDS shape. \cite{tm97} further developed
a more realistic SOC model to include the  dynamic processes
of  advection-dominated accretion disks. 
The viscosity parameter $\alpha$ is switched to a higher value when
the surface density exceeds a critical value. Under this prescribed
critical condition and the assumption of a uniform disk temperature, they showed 
that the improved SOC model can generate  light curves and PDS
similar to the observed ones.  For a  uniform temperature disk, the time profile of 
the thermal bremsstrahlung emissions is energy independent, and thus
the PDS shape is also energy independent. To explain the energy dependent 
effects we have shown here, it will be necessary to expand this SOC model 
to the case of a multi-temperature disk. Such a multi-temperature
disk is also needed to explain the X-ray energy spectra that 
have power-law photon indices higher than 1. 

Like the SOC model, the wave propagation model (\cite{manmoto96}) 
also needs an assumption for the temperature profile in the disk. The wave 
is generated by disturbances such as magnetic flares at the outer part 
of the disk and propagates into the event horizon without much damping. 
The wave model can account for the time profiles of individual  X-ray shots, 
and predicts that the energy of the emitted photons increases as the 
wave propagates into the horizon because the local temperature 
of the disk increases with decreasing radius. This can explain the hard
lags of the X-ray emission. The time profile of the X-ray flux in each 
energy band is determined by the velocity of the wave and the 
temperature profile of the disk. If the radial velocity is constant and the
temperature distribution is linear along the radial direction, the time 
profiles of the shots are the same for all the energy bands, and 
then the PDS shape should be energy independent. This is consistent 
with our analysis results of GRS 1758--258, 1E 1740.7--2942, and 
GX 339--4. A more realistic version of the model will be required to 
explain the energy dependent behaviors of Cygnus X--1.

\subsection{Comptonization Models}

The hard-low state spectrum of GBHCs can best be interpreted as the
unsaturated Comptonization of soft photons by hot thermal electrons
(\cite{katz76}; \cite{liang98}). However, measurements of the Fourier-frequency 
dependence of the time lags between the signals in different X-ray energy 
channels indicates that this pure Comptonization model  is not 
applicable to a moderate-sized uniform plasma (\cite{miyamoto88}).
\cite{kazanas97} and \cite{hua97} found that Comptonization 
in a hot, inhomogeneous corona with a density profile $n(r) \propto 
r^{-1}$ can fit the combined spectral and timing 
properties of Cygnus X-1, although the hot corona has to 
extend out to $\sim 1$ light second from the black hole. 
B\"ottcher \& Liang (1998) studied the isolated 
effects of Comptonization on the rapid aperiodic variability of 
black hole X-ray binaries in great detail, investigating different 
source geometries. In the case of
a central soft photon source, corresponding to an accretion-disk
corona geometry with a covering fraction of the corona close to
unity, the PDS was generally found to become steeper for increasing
photon energies. This trend becomes weaker as the density 
gradient of the hot, thermal corona becomes steeper. If the soft
photon source is external to the hot corona, which corresponds
to an ADAF-like geometry (\cite{ny94}; \cite{chen95}; \cite{emn97}; \cite{ll98})
with a cool outer accretion disk or a radiation-pressure 
dominated, hot inner accretion disk solution (\cite{sle76}), only a 
weak steepening of the PDS is predicted. This steepening
trend is due to the averaging effect of Compton scatterings.
Low energy photons, which have not gone through many up-scatterings,  
keep most of the variation of the soft photon input, while the 
high energy photons, having been accumulated in the 
corona for a longer time $t$,  reflect the average of 
the soft photon input over a time $t$, and thus become
less variable at high frequencies.  For the same reason, the overall rms 
amplitude of the variability is expected to decrease with increasing photon 
energy. These predictions are not consistent with the data
presented here. No cases of a steepening PDS shape were found. Therefore, 
the pure Compton scattering model does not fully explain the 
emission mechanism of GBHCs even for  an inhomogeneous corona.

To overcome the problems with the pure Comptonization model, 
B\"ottcher \& Liang (1999) developed a drifting blob model. 
It was motivated by the inhomogeneous accretion 
equilibrium found by \cite{krolik98} and assumes that dense blobs 
of cool material are emitting soft photons and spiraling inward 
through a hot, optically thin inner accretion flow, which could be an ADAF or a 
Shapiro-Lightman-Eardley type hot inner disk. In this model, 
the hard time lags are due to the radial drifting time that the dense blobs
spend in traveling from the cool to hot parts of the accretion flow.
The model does not require a big Comptonization corona to 
account for the hard time lags, and thus the photon diffusion time
is negligible. The photon flux resulting from a single drifting blob
has significantly different time profiles in different energy bands.
The low energy photon flux persists for the whole drifting period
because most of the soft photons being steadily emitted by the drifting 
blob emerge from the corona without being uperscattered to higher energy.
On the other hand, most of the high energy photons emerge only at later times 
when the blob drifts to the inner hot part of the disk. Therefore, the
high energy shots are much narrower than the low energy shots, and thus
have a broader band in frequency space. Consequently, the PDS is 
predicted to become flatter with increasing photon energy. This is 
consistent with the observations of Cygnus X--1, which has  flatter 
power spectra at higher energies. In order to reproduce the complete 
PDS of individual objects, additional assumptions about the production of these
inward-drifting blobs are required. In the simplest picture, the overall rms 
amplitude of the variability is expected to increase  with increasing 
photon energy. Since we have shown here that this is not always the 
case, a more detailed model may be required.

Another Compton scattering related model is the magnetic flare model proposed by 
\cite{pf99}, which involves active regions of a patchy corona above an 
optically thick accretion disk. Driven by the hydrodynamic and radiation pressure, 
the active regions are moving away from the disk and being heated up 
by the magnetic energy.  As the active regions move outward, they 
become hotter because there is less Compton cooling.
The feedback radiation from the active regions also heats up
the disk regions that provide the soft photons for the Compton
scatterings in the active regions. These heating processes make 
the observed photon energy increase as the flares proceed, and 
thus the low energy photon flux reaches a maximum earlier than the high energy flux. 
The light curves at different energies are self-similar but with
higher characteristic time scales for higher energy photons. Consequently,
the power density spectra in different energies have similar shapes with a
 energy independent $\beta$ and a decreasing $f_2$ for higher energies. 
The first broken power law is determined by the avalanche processes,
and thus it is energy independent. The PDS ratios would then decrease 
at frequencies above $f_1$. However, we do not see this effect 
in any of our observations, implying that this version
of the magnetic flare model is not supported.

\section{Discussion}

Our analysis shows that the classical black hole candidate Cygnus 
X--1 has significantly different temporal properties from 1E 1740.7-2942, 
GRS 1758--258 and GX 339-4 in spite of their similarities in energy spectra. 
The different energy dependence of the PDS shape may indicate  that 
the emission mechanism in Cygnus X--1 is different from the other three 
GBHCs. If this is true, it will be necessary to develop
accretion disk models that can accommodate this difference.
The spiraling blob model, for example, could be the dominant mechanism
for Cygnus X-1 while other mechanisms such as the wave propagation model
may be more appropriate for the other three sources. However,
these models do not exclude each other. Each of the models may 
just describe one aspect of the emission process.
The SOC and wave models, for example, can be the blob generation mechanisms
for the drifting blob model. Combining these models together will certainly
make them more realistic and robust.

The analysis of the overall variability is complicated by the fact
that the X-ray light curves may consist of two components: the shot component
and the persistent component (Takeuchi \& Mineshige 1997), which may have different
origins and spectra. It is difficult to separate these two components
spectrally and temporally. The integrated rms amplitude 
reflects total variability of the source emission and it is hard to directly compare
the models that usually only simulate the shot processes.
The SOC model is an exception. It generates shots using processes that occur
in the steady accretion flow and,  therefore, links the two components together.
However, the energy dependence of the variability is still an open
question for the SOC model. 

The reason for  the anti-correlating trend found between
the integrated rms amplitude and the X-ray flux may also lie in the
superposition of the shot and persistent components. If the radiation, for example,
is dominated by the persistent component, any increase in the persistent
flux would increase the total flux and thus reduce the relative variability, i.e.,
the integrated rms  amplitude. If the emitting regions of the two components
are physically separate, and there are locally more absorption materials
around the shot emitting regions, then a high absorption would mean less variability.
This would explain the anti-correlating trend between the rms amplitude and
the hydrogen column density.

\acknowledgements{
This work was supported by NASA grant  NAG 5-3824 at Rice University.
We thank Wei Cui and Juri Poutanen for their helpful discussions, David 
Smith for his assistance with the background determination for 1E 1740.7--2942
and GRS 1758--258, Jim Lochner and Gail Rohrbach of the $RXTE$ team for their
help with the data processing, and the anonymous referee for invaluable 
comments and suggestions. The work of M.B. is supported by NASA 
through Chandra Postdoctoral Fellowship Grant number PF9-10007 
awarded by the Chandra X-ray Center, which is operated by the 
Smithsonian Astrophysical Observatory for NASA under
contract NAS8-39073.}

\clearpage

\begin{deluxetable}{lllll}
\tablewidth{16.5cm}
\tablecaption{ The observation date of each curve in Figure 5 and its related
{\it RXTE} ASM (2 -- 12 keV) daily count rate.}
\tablehead{
\colhead{\#\tablenotemark{a}} & \colhead{Cyg X--1} &  \colhead{GX 339--4} &  
\colhead{GRS 1758--258} &  \colhead{1E 1740.7--2942} }

\startdata

\begin{tabular}{l}
\\ 1 \\ 2 \\ 
 \\ 3 \\ 4 \\ 5 \\ 6 \\7
\end{tabular}

&
\begin{tabular}{ll}
Date \tablenotemark{b}  & ASM \tablenotemark{c}\\
465  & $19.7 \pm 0.7$  \\
467  & $17.7 \pm 0.4$  \\
& \\
725  & $26.0 \pm 0.4$  \\
128  & $34.3^{*}$         \\
125  & $33.8^{*}$       \\
130  & $34.6^{*}$        \\
624  & $30.0 \pm 0.4$ \\
\end{tabular}

&
\begin{tabular}{ll}
Date \tablenotemark{b}  & ASM \tablenotemark{c}\\
636 & $1.4 \pm 0.5$   \\
749 & $1.4 \pm 0.7$     \\
& \\
597 & $1.7 \pm 0.5 $  \\
711 &  $4.0 \pm 0.5$   \\
482 &  $3.1 \pm 0.8$  \\
553 &  $5.5 \pm 0.3$  \\
& \\
\end{tabular}

&
\begin{tabular}{ll}
Date \tablenotemark{b}  & ASM \tablenotemark{c}\\
646   & $1.4 \pm 0.5 $  \\
673/74/78 & $1.7 \pm 0.5 $  \\
/80/84 & \\
562                   & $1.7 \pm 0.4  $  \\
513                   & $3.1 \pm 0.8  $ \\
470                   & $7.6   \pm 2.0$ \\
517                   & $6.3 \pm 1.3  $ \\
& \\
\end{tabular}

&
\begin{tabular}{ll}
Date \tablenotemark{b}  & ASM \tablenotemark{c}\\
504          & $1.7 \pm 0.4$ \\
158--61  & $2.9 \pm 0.6$  \\
& \\
764          & $2.8 \pm 1.5$ \\
763          & $2.1^{*}$  \\
584--88  & $2.4 \pm 0.3$\\
& \\
& \\
\end{tabular}

\enddata
\tablenotetext{a}{The curve number 
(the first column) is counted from top to bottom in the medium 
energy band (5 -- 10 keV) in Figure 5; thus the rms amplitudes are in the 
decreasing order for each source. }
\tablenotetext{b}{The date is the Modified Julian Date  -- 50000.}
\tablenotetext{c}{The count rate is in counts/s, and 
the count rates superscribed by `*'  are interpolated from the measurements 
before and after the observation date.}

\end{deluxetable}

\begin{figure}[p]
\begin{center}
\epsfysize=12cm
\epsffile{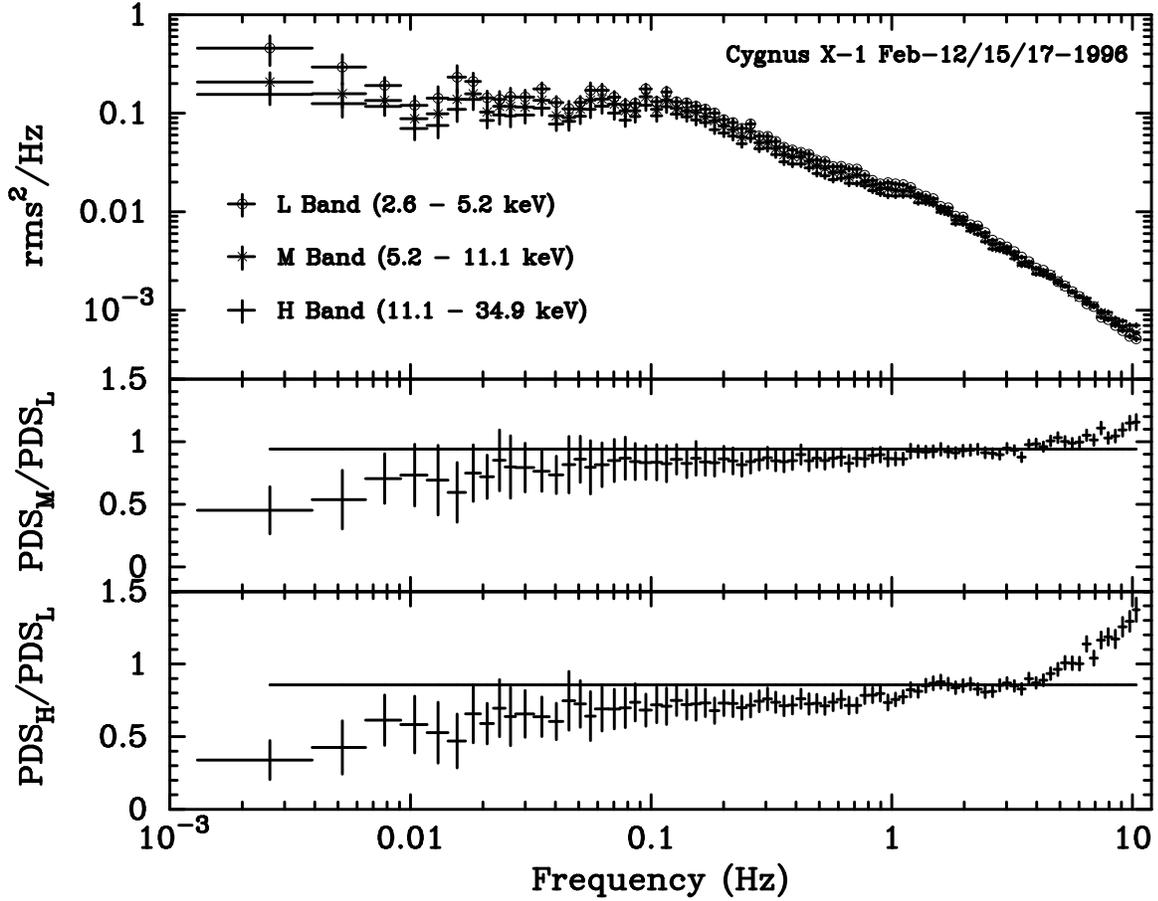}
\end{center}
\caption{A sample PDS of Cygnus X-1 and the PDS ratios between 
the energy bands. The upper panel is the PDS for the three energy 
bands. The lower two panels are the PDS ratios between the L 
band and the other two bands. Constant fits to the PDS ratios 
over the whole frequency range give a $\chi^{2}_{\nu} $ = 1.85 
for 82 degrees of freedom (DOF) and null hypothesis
probability Q = $4.47 \times 10^{-6}$ for ${\rm PDS_{M}/PDS_{L} }$
and a $\chi^{2}_{\nu} $ = 5.6 (DOF = 82) and Q = $7.2 \times 10^{-54}$ 
for ${\rm PDS_{H}/PDS_{L} }$. The increasing PDS ratio indicates 
a flatter power spectrum 
at higher energies.}
\end{figure}

\begin{figure}[p]
\begin{center}
\epsfysize=12cm
\epsffile{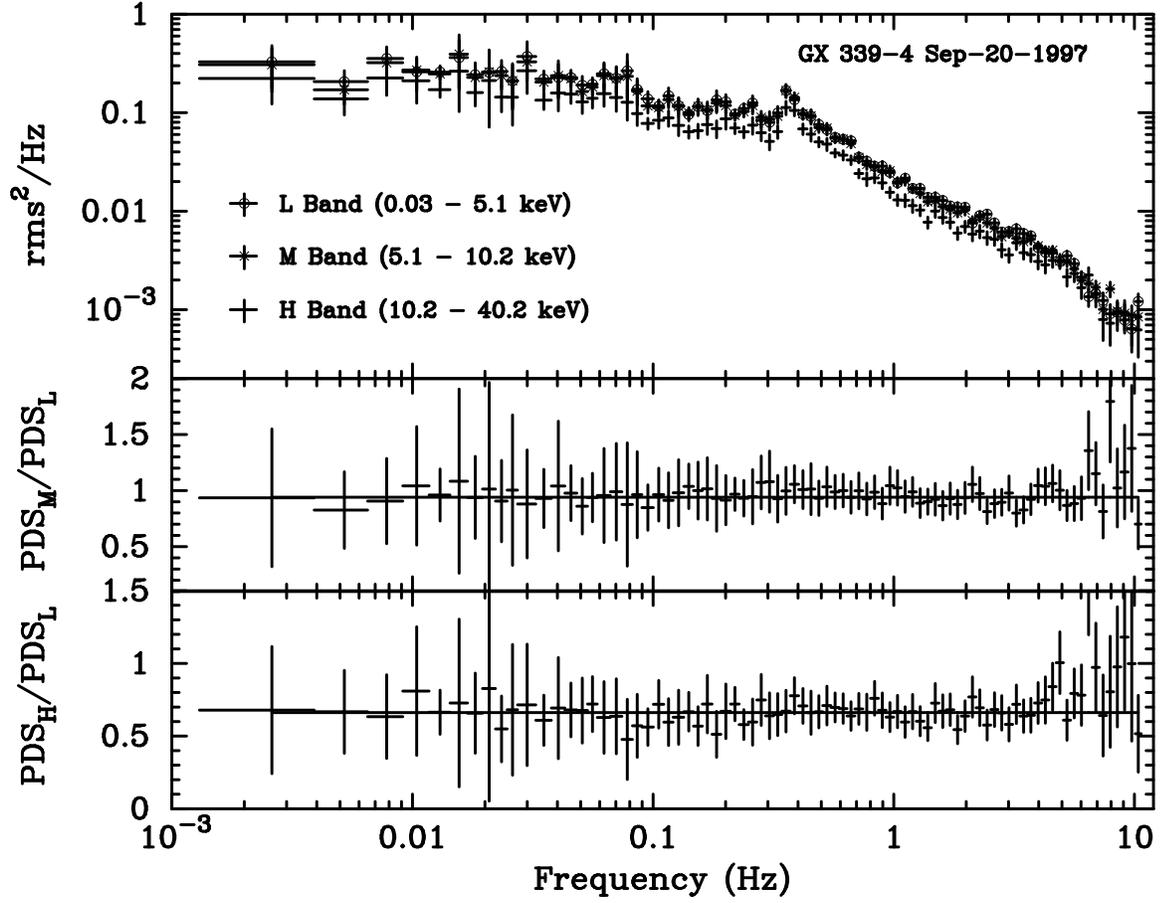}
\end{center}
\caption{A sample PDS of GX 339-4 and the PDS ratios between the
energy bands. Constant fits to the PDS ratios over the whole 
frequency range give a $\chi^{2}_{\nu} $ = 0.23 (DOF = 82) 
and Q $\approx$ 1.0 for ${\rm PDS_{M}/PDS_{L} }$ and a 
$\chi^{2}_{\nu} $ = 0.41 (DOF = 82)
and Q  $\approx$ 1.0  for ${\rm PDS_{H}/PDS_{L} }$. The flat PDS
ratio indicates an energy independent shape of the power
spectrum.}
\end{figure}

\begin{figure}[p]
\begin{center}
\epsfysize=12cm
\epsffile{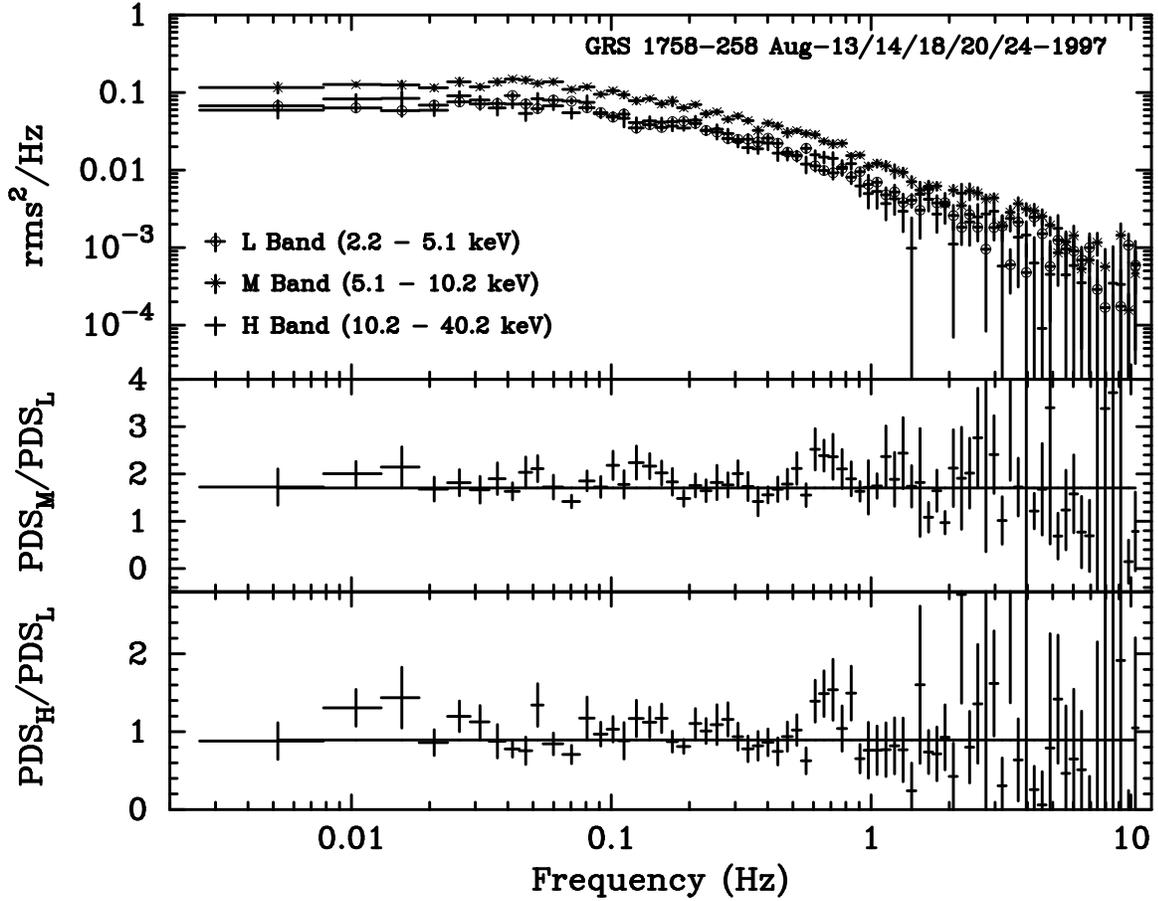}
\end{center}
\caption{A sample PDS of GRS 1758-258 and the PDS ratios between 
the energy bands. Constant fits to the PDS ratios over the whole 
frequency range give a $\chi^{2}_{\nu} $ = 1.17 (DOF = 72) 
and Q = 0.15 for ${\rm PDS_{M}/PDS_{L} }$
and a $\chi^{2}_{\nu} $ = 0.41 (DOF = 72) and Q = 0.24 for
${\rm PDS_{H}/PDS_{L} }$. }
\end{figure}

\begin{figure}[p]
\begin{center}
\epsfysize=12cm
\epsffile{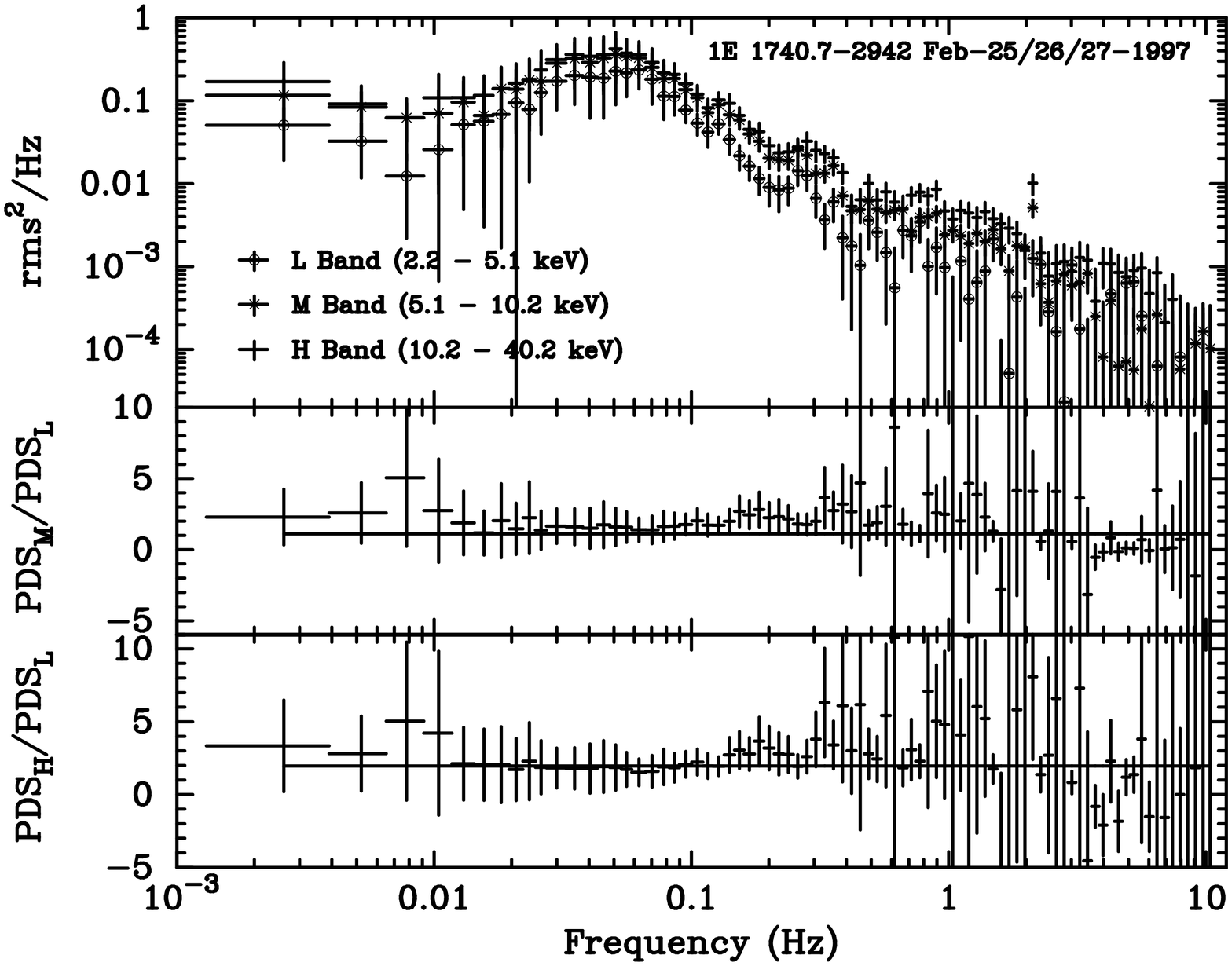}
\end{center}
\caption{A sample PDS of 1E 1740.7-2942 and the PDS ratios between the
energy bands. Constant fits to the PDS ratios over the whole frequency 
range give a $\chi^{2}_{\nu} $ = 0.94 (DOF = 82) and Q = 0.76 for 
${\rm PDS_{M}/PDS_{L} }$ and a $\chi^{2}_{\nu} $ = 0.41 (DOF = 82) 
and Q $\approx$ 1.0 for ${\rm PDS_{H}/PDS_{L} }$.}
\end{figure}

\begin{figure}[p]
\begin{center}
\epsfysize=12cm
\epsffile{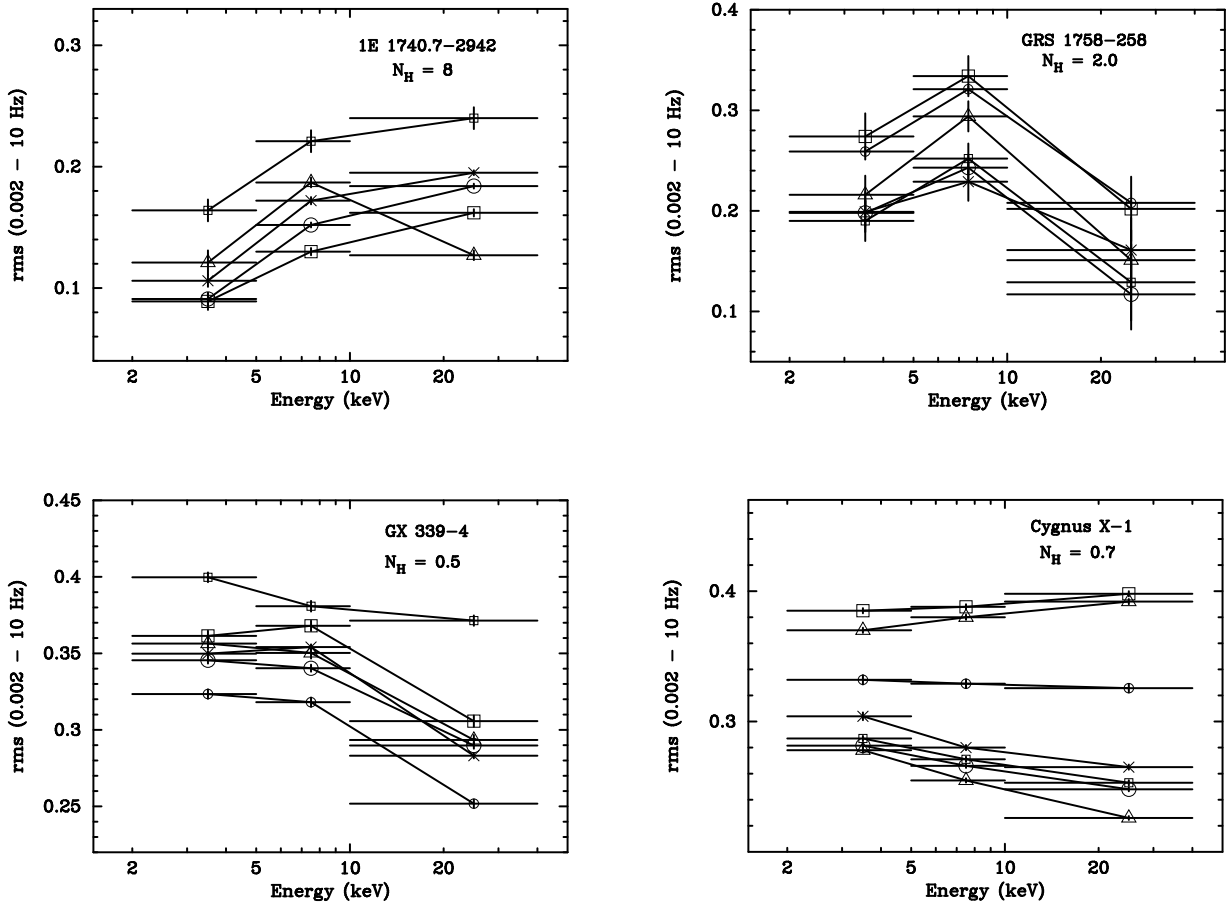}
\end{center}
\caption{The overall variability vs energy. The rms is calculated by
integrating the PDS over the frequency range of 0.002 -- 10~Hz and
taking the square root of the integral. Each connected curve
represents one observation with the corresponding dates and the 
related ASM daily count rates listed in Table 1.  $N_H$ gives the 
typical hydrogen column density measured for the four sources in 
units of $10^{22}$~cm$^{-2}$ (1E 1740.7--2942: Sheth et al. 1996; 
GRS 1758--258: Lin et al. 1999; GX 339--4: M\'endez \& Van der 
Klis 1997; Cyg X--1: Baluci\'nska-Church et al. 1995). 
The energy dependence of the overall variability is
significantly different among the four sources.}
\end{figure}

\begin{figure}[p]
\begin{center}
\epsfysize=12cm
\epsffile{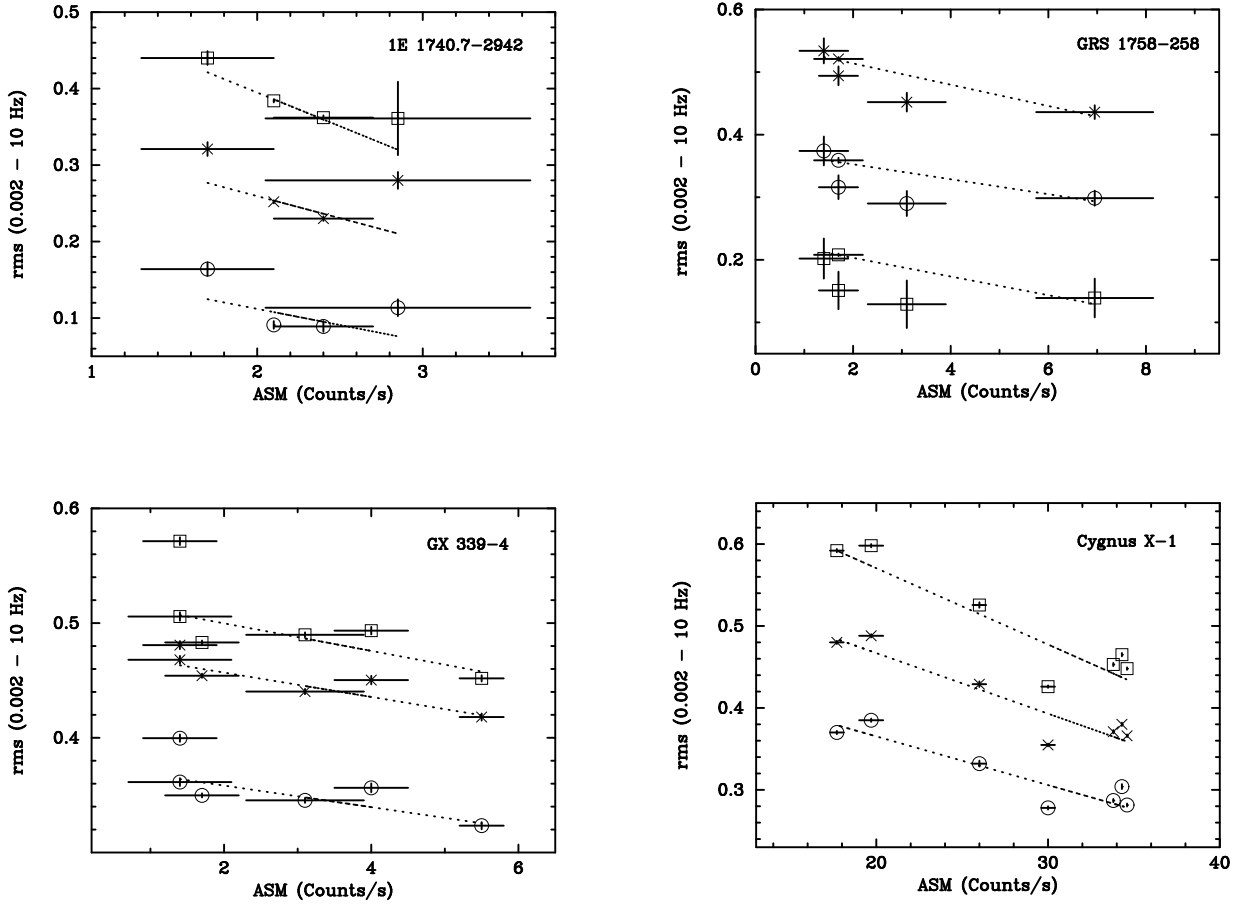}
\end{center}
\caption{The anti-correlation between the integrated rms amplitude 
and count rate of $RXTE$ ASM (2 -- 12 keV). For illustrative purposes,
the rms amplitudes for band M (crosses)  have been  shifted upwards 
by 0.1, and band H (squares) by 0.2 except for GRS 1758-258, whose 
rms amplitude in band H has been shifted upwards by 0.3. No shifts 
have been made to band L (circles). The dotted lines are the best 
fit linear functions for each data set.}
\end{figure}


\clearpage

\begin{thebibliography} {}

\bibitem[Baluci\'nska-Church et al. 1995]{b95}
Baluci\'nska-Church, M., Belloni, T., Church, M. J., \& Hasinger, C. 1995, A\&A, 302, L5

\bibitem[B\"ottcher \& Liang 1998] {bl98}
B\"ottcher, M., \& Liang, E. P. 1998, ApJ, 506, 281

\bibitem[B\"ottcher \& Liang 1999] {bl99}
B\"ottcher, M., \& Liang, E. P. 1999, ApJ, 511, L37 

\bibitem[Chen et al. 1995]{chen95}
Chen, X., Abramowicz, M. A., Lasota, J.-P., Narayan, R., \& Yi, I. 1995, ApJ, 443, L61

\bibitem[Cui et al. 1997] {cui97}
Cui, W., Zhang, S. N., Focke, W., \& Swank, J. H. 1997, ApJ,
484, 383

\bibitem[Cui 1998] {cui98}
Cui, W. 1998, in ``High-Energy Processes in Accreting Black
Holes'', eds. J. Poutanen \& R. Svensson, ASP Conf. Series, Vol. 161,
p. 97

\bibitem[Esin, McClintock \& Narayan 1997]{emn97}
Esin, A. E., McClintock, J. E., \& Narayan, R. 1997, ApJ, 489, 865

\bibitem[Hua et al. (1997)]{hua97}
Hua, X.-M., Kazanas, D., \& Titarchuk, L. 1997, ApJ, 482, L57

\bibitem[Katz 1976]{katz76}
Katz, J. I. 1976, ApJ, 206, 910

\bibitem[Kazanas et al. (1997)]{kazanas97}
Kazanas, D., Hua, X.-M., \& Titarchuk, L. 1997, ApJ, 480, 735

\bibitem[Krolik (1998)]{krolik98}
Krolik, J. H. 1998, ApJ, 498, L13

\bibitem[Liang 1998]{liang98}
Liang, E. 1998, Phys. Rep., 302, 67

\bibitem[Lin et al. 1999] {lin99}
Lin, D.,  Smith I. A., Liang, E. P.,  Bridgman, T., Smith, D. M., Mart\'{\i}, J.,  
Durouchoux, Ph., Mirabel, I. F.,  \& Rodr\'{\i}guez, L. F. 1999, ApJ, submitted

\bibitem[Lin et al. 2000] {lin00}
Lin, D.,  Smith I. A., Liang, E. P.,  B\"ottcher, M. 2000, in preparation

\bibitem[Luo \& Liang 1998]{ll98} 
Luo, C., \& Liang, E. P. 1998, ApJ, 498, 307

\bibitem[Maejima et al. 1984]{maejima84}
Maejima, Y., Makishima, K., Matsuoka, M., Ogawara, Y., Oda, M., 
Tawara, Y., \& Doi, K. 1984, ApJ, 285, 712 

\bibitem[Main et al (1999)]{main99}
Main, D. S., Smith D. M., Heindl W. A. , Swank J., Leventhal M., 
Mirabel, I. F., Rodr\'{\i}guez, L. F. 1999, to appear in ApJ, astro-ph/9906178

\bibitem[Manmoto et al. 1996]{manmoto96}
Manmoto, T., Takeuchi, M., Mineshige, S., Matsumoto, R., 
\& Negoro, H. 1996, ApJ, 464, L135

\bibitem[M\'endez \& Van der Klis 1997]{mv97}
M\'endez, M., \& Van der Klis, M. 1997, ApJ, 479, 926

\bibitem[Miyamoto et al. 1988, 1993]{miyamoto88}
Miyamoto, S., Kitamoto, S., Mitsuda, K., \& Dotani, T. 1988, Nature, 336, 450

\bibitem[Miyamoto et al. 1992]{miyamoto92}
Miyamoto, S., Kitamoto, S., Iga, S., Negoro, H., \& Terada, K. 1992, ApJ, 391, L21 

\bibitem[Miyamoto et al. 1988, 1993]{miyamoto93}
Miyamoto, S., Iga, S., Kitamoto, S., \& Kamado, Y. 1993, ApJ, 403, L39 

\bibitem[Narayan \& Yi 1994]{ny94}
Narayan, R., \& Yi, I. 1994, ApJ, 428, L13 

\bibitem[Nolan et al. 1981]{nolan81}
Nolan, P. L., et al. 1981, ApJ, 246, 494

\bibitem[Nowak et al. 1998]{nowak98}
Nowak, M. A., Vaughan, B. A., Wilms, J., Dove, J. B., \&Begelman, 
M. C. 1998, ApJ, 510, 874

\bibitem[Oda et al. (1971)]{oda71}
Oda, M., Gorenstein, P., Gursky, H., Kellogg, E., Schreier, E., Tananbaum, H., 
\& Giacconi, R. 1971, ApJ, 166, L1


\bibitem[Poutanen \& Fabian (1999)] {pf99}
Poutanen, J., \& Fabian, A. C. 1999, MNRAS, 306, L31

\bibitem[Shapiro, Lightman \& Eardley 1976]{sle76}
Shapiro, S., Lightman, A. P., \& Eardley, D. M. 1976, ApJ, 204, 187 

\bibitem[Sheth et al. 1996]{s96}
Sheth, S., Liang, E., Lou, C., \& Murakami, T. 1996, ApJ, 468, 755


\bibitem[Smith \& Liang (1999)]{sl99}
Smith, I. A., \& Liang, E. P. 1999, ApJ, 519, 771

\bibitem[Strohmayer et al. 1997]{strohmayer97}
Strohmayer, T. E., Jahoda, K., Giles, A. B., \& Lee, U. 1997, ApJ, 486, 355


\bibitem[Takeuchi, Mineshige \& Negor (1995)]{takeuchi95}
Takeuchi, M., Mineshige, S., \& Negoro, H. 1995, PASJ, 47, 617

\bibitem[Takeuchi \& Mineshige (1997)]{tm97}
Takeuchi, M., \& Mineshige, S. 1997, ApJ, 486, 160

\bibitem[Terrell (1972)]{terrell72}
Terrell, N. J. 1972, ApJ, 174, L35

\bibitem[van der Klis 1995]{van95}
van der Klis, M. 1995, in X-Ray Binaries (Cambridge: Cambridge Univ. Press)


\end{thebibliography}
\end{document}